\documentstyle[12pt,epsfig]{iopart}
\def\las{\mathrel{\hbox{\rlap{\hbox{\lower4pt\hbox{$\sim$}}}\hbox{$<$}}}} 
\def\gas{\mathrel{\hbox{\rlap{\hbox{\lower4pt\hbox{$\sim$}}}\hbox{$>$}}}}

\def\ca19{Ca{\sc\,xix}}
\def\etal{{\it et \thinspace al.}\ } 
\begin{document} 
\jl{2} 
\title[EIE He-like ions]{Electron Impact Excitation of Helium-like 
ions up to n=4 levels including radiation damping} 
 
\author{Franck Delahaye$^{\dag,\ddag}$, Anil K. Pradhan$^{\dag}$ and Claude J. 
Zeippen$^{\ddag}$} 
 
\address{\dag Department of Astronomy,  
Ohio State University \\ Columbus, Ohio, 
USA, 43210}
\address{\ddag LUTH (UMR 8102 associ\'ee au CNRS et \`a l'Universit\'e Paris
7),\\ Observatoire de Paris, F-92195 Meudon, France } 
 
\begin{abstract} 
 
Helium-like ions provide the most important X-ray spectral diagnostics
in high temperature fusion and astrophysical plasmas.
We previously presented computed collision strengths
for O~VII including relativistic fine structure, levels up to the $n=4$
complex and radiation damping of autoionizing resonances. We have extended
this work to other He-like ions (N, Ne, Mg, Al, Si, S, Ca).
The calculations are carried out using the Breit-Pauli R-matrix (BPRM) method 
with a 31-level eigenfunction expansion.  
Collision strengths for the principal lines important in X-ray plasma 
diagnostics, w, x, y and z, corresponding to the 4 transitions to
the ground level $1s^2 \ (^1S_0)
\longleftarrow 1s2p (^1P^o_1), 1s2p (^3P^o_2), 1s2p (^3P^o_1), 1s2s
(^3S_1)$, are explicitly shown. We find the effect of radiation damping  
to be significant for the forbidden transitions in heavier He-like ions, 
which should affect the diagnostic line ratios.
We extrapolated the collision strengths to their values at infinite energy  
using
the Burgess-Tully extrapolation technique. This is required to calculate
the Maxwellian average collision strengths at high temperature. 
We show that the coupling between dipole allowed and inter-combination
transitions affects increasingly the effective collision strengths for the
$n ^1S_0 - n\prime ^3P_1$ transition as the charge of the ion increases. This 
clearly affects the treatment of the extrapolation toward the infinite energy
point of the collision strength.
This work is carried out as part of the Iron Project-RmaX Network. 

\end{abstract}

\pacno{34.80.Kw} 
\maketitle 
 
\submitted 
 
\section{Introduction} 
  
Helium-like ions provide the most important X-ray spectral diagnostics
in high temperature fusion and astrophysical plasmas.
The new generation of X-Ray satellites such as the Chandra X-Ray
Observatory and the X-Ray Multi-Mirror Mission-Newton provide
high resolution spectra of
different types of astronomical objects (e.g. Kaastra \etal 2000,
Porquet and Dubau 2000, Porquet \etal 2001). The high sensitivity of these 
observatories and the high quality of the spectra they produce requires
highly accurate atomic data for a precise interpretation. 
The aim of  the Iron Project-R-matrix calculations for X-ray  
spectroscopy (IP-RmaX) is to calculate extended sets of accurate collision  
strengths and rate coefficients for all ions of importance in X-Ray  
diagnostics. 
Among previous works, the electron impact excitation of
Helium-like ions was considered by Sampson \etal (1983)
and Zhang \& Sampson (1987). They used the Coulomb-Born approximation with
exchange, intermediate coupling, and some resonance effects to  
obtain collision strengths for Helium-like ions, with atomic number
Z spanning a large  
range of values ($4<Z<74$). The present work aims at generating a more complete 
dataset of  
high reliability for He-like ions, including all important effects for 
highly-charged ions such as relativistic effects, radiation damping, and  
resonances in higher complexes up to $n=4$.

The method and computations are summarized in section 2.
Results for the collision strengths and important issues  
are discussed in section 3, and the present results for the effective
(Maxwellian averaged) collision strengths are
compared with previous calculations. The main conclusions are given in
section 4.

\section{Method and Computations} 
 
The collisional calculation in the present work has been carried out using 
the Breit-Pauli R-matrix (BPRM) method as used in the Iron Project (IP) and
utilized in a number of previous publications.
The aims and methods of the IP are presented in Hummer \etal (1993).
We briefly summarize the main features of the method and present calculations.

In the coupled channel or close coupling (CC) approximation
the wave function expansion,
$\Psi(E)$, for a total spin and angular symmetry  $SL\pi$ or $J\pi$,
of the (N+1) electron system
is represented in terms of the target ion states as:

\begin{equation}
\Psi(E) = A \sum_{i} \chi_{i}\theta_{i} + \sum_{j} c_{j} \Phi_{j},
\end{equation}

\noindent
where $\chi_{i}$ is the target ion wave function in a specific state
$S_iL_i\pi_i$ or level $J_i\pi_i$, and $\theta_{i}$ is the wave function
for
the (N+1)$^{th}$ electron in a channel labeled 
$S_iL_i(J_i)\pi_i \ k_{i}^{2}\ell_i(SL\pi) \ [J\pi]$; $k_{i}^{2}$ is the
incident kinetic energy. In the second sum the $\Phi_j$'s are
correlation
wave functions of the (N+1) electron-system that (a) compensate for the
orthogonality conditions between the continuum and the bound orbitals,
and (b) represent additional short-range correlations that are often of
crucial importance in scattering and radiative CC calculations for each
symmetry. The $\Phi_j$'s are also referred to as ``bound channels", as
opposed to the continuum or ``free" channels in the first sum over the
target states. In the relativistic BPRM calculations the set of
${SL\pi}$
are recoupled in an intermediate (pair) coupling scheme 
to obtain (e + ion) states  with total $J\pi$, followed by
diagonalization of the (N+1)-electron Hamiltonian. Details of the
diagonalization and the R-matrix method are given in many previous works
(e.g. Berrington \etal 1995).

The target expansion for the close coupling calculations consists of 31  
fine-structure levels arising from the 19 LS terms with principal quantum  
number $n\leq 4$. The target eigenfunctions were developed using the
SUPERSTRUCTURE program
(Eissner \etal 1974) in a version due to Nussbaumer and Storey (1978).
The full expansion consists in 10 configurations up to n=4:
$1s^2, 1s2s, 1s2p, 1s3s, 1s3p, 1s3d, 1s4s, 1s4p, 1s4d$ and $1s4f$
The scaling factors have been obtained by first optimizing the $\lambda_{1s}$
parameter through the minimization of the energy of the ground term. Then 
all the other scaling parameters have been optimized on the 19 terms
considered here.
The scaling  factors in the {\it Thomas-Fermi} potential employed in 
SUPERSTRUCTURE can be obtained upon request from the first author.
The optimization of the orbitals has been performed as follow:
 
In order to estimate the quality of the target wavefunction expansion,  
we compare the energy levels with those from the {\it National Institute
for Standards and Technology} (NIST). The agreement with the NIST values is 
found to be very good, within 0.05\% for all levels. A better
criterion for the accuracy of the wavefunctions is the accuracy
of the oscillator strengths for transitions in the target ion .
The oscillator strengths agree well within 10\% (however, some 
of the values are given by NIST with an estimated accuracy  
of 30\%). Another accuracy criterion is the level of agreement between the 
oscillator strengths in the length and the velocity formulations,
which we also find to be a few percent for all transitions. 

The 19 LS terms are recoupled in the relativistic BPRM calculations 
into the corresponding 31 fine-structure 
levels up to the  $n=4$ complex using the routine
RECUPD that performs intermediate coupling
operations including the one-body Breit-Pauli operators (Hummer et. al.
1993). The reconstructed target eigenfunctions and the
resulting target energies reproduce to $10^{-4}$ Ryd the results from  
SUPERSTRUCTURE, verifying that the algebraic operations have been
carried out self-consistently and without loss of accuracy. 
The collision strengths have been calculated for electron  
energies $0 \leq E \leq 4 \times E(4 ^1P_1)$ Ryd. This wide energy range 
ensures a good coverage of the region where resonances up to the $n=4$ complex 
are important, as well as the higher energy region where no resonance has 
been included (all channels are open) but where the background collision 
strengths still make a significant contribution to the Maxwellian averaged 
rate coefficient for electron temperatures of interest. However, it might not
be sufficient for the calculation of collision rates at high temperature. Given
the slow convergence of some transitions and the need for high partial wave at
high temperature, it is unpractical to carry the calculations at higher
energy. However it is possible to extrapolate the collision strengths to the
value at infinite electron energy using the procedure from Burgess and Tully
(1993). 
 
The inner region R-matrix basis set included 40 orbitals per angular
momentum.
Because of the importance of the near threshold resonances in the Maxwellian  
average rate coefficient, careful attention has been devoted to the  
resolution and a precise mesh has been chosen. A mesh of $10^{-5}$ Ryd was  
selected for the region where resonances are important, and a coarser mesh for 
the region where all channels are open. 
We included the contribution to the collision strengths from all symmetries
with total angular momentum J and both odd and even parities,  
$J\pi \leq (\frac{35}{2})^{o,e}$. The contribution of higher partial waves
 was included  
using the Coulomb-Bethe approximation via the `top-up' facility 
in the asymptotic region program STGF of the R-matrix package
(Burke and Seaton 1986; modified by W. Eissner and G.X. Chen).

\section{Results and discussion\protect\\} 

In figures 1 and 2 we present the collision strengths for transitions from  
$2 ^3S_0$ and $2 ^1P^o_1$ respectively to the ground state for all 
ions included in the present calculation. The high resolution of the calculations with
a large number of points allows us to resolve clearly all the resonances up to
the last threshold in the $n=4$ complex. We delineate the Rydberg series
converging to the different series limits in all three complexes.
The doubly excited (e + He-like ion ) $\rightarrow$ Li-like ion resonance complexes, 
KMM, KMN etc.. converging  
towards the different  $n=3$ and $n=4$ levels are clearly resolved. 
 
%---------------------------------------------------------------------------- 
\begin{figure} 
%\vspace*{-2.0cm} 
\centering 
\psfig{figure=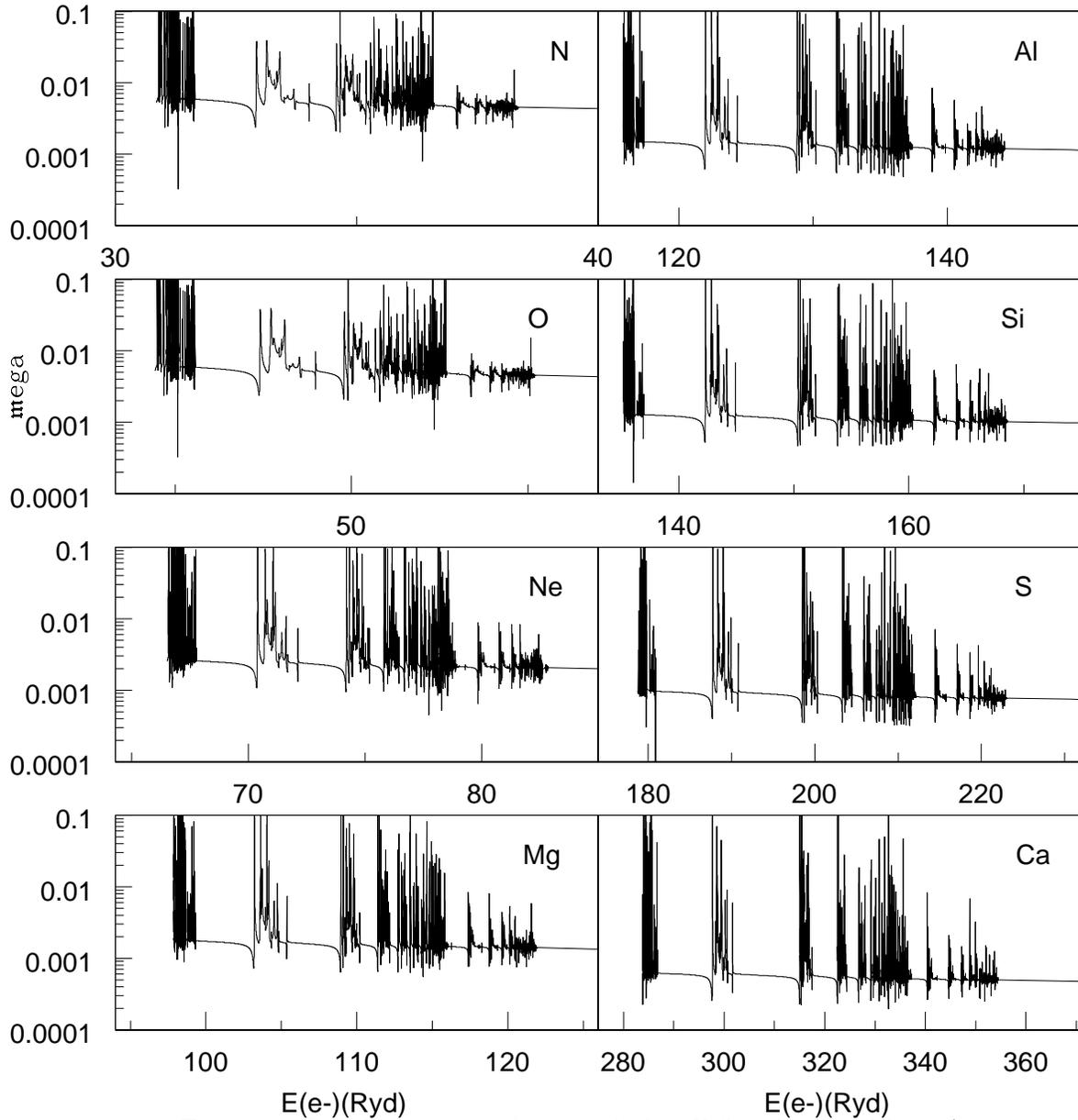,height=17.0cm,width=17.0cm} 
\vspace*{-1cm} 
\caption{Collision strengths for the principal X-Ray line transition z 
(from ground state $1s^2$ $^1S_0$ to $2 ^3S_1$).} 
\end{figure} 
 
\begin{figure} 
%\vspace*{-2.0cm} 
\centering 
\psfig{figure=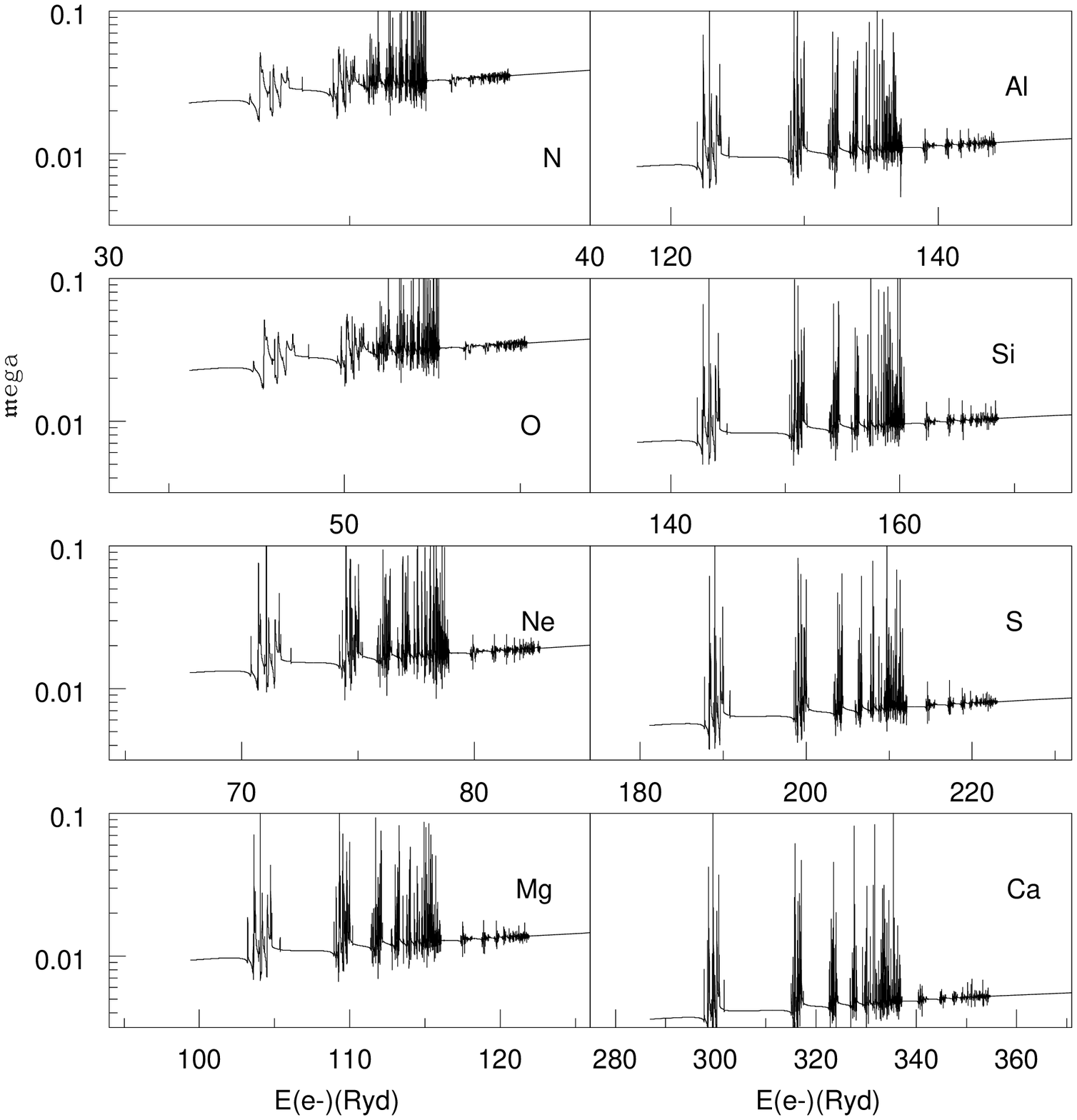,height=17.0cm,width=17.0cm} 
\vspace*{-1cm} 
\caption{ Collision strengths for transitions corresponding to the 
w (or r) line ($1s^2\ ^1S_0 \rightarrow  2 ^1P^o_1$ )} 
\end{figure}

%---------------------------------------------------------------------------- 
\subsection{Radiation Damping} 

It has been previously shown (Presnyakov and Urnov 1979, Pradhan 1981,
Pradhan and Seaton 1985), that radiation damping may have a significant 
effect on the resonances in collision strengths for highly charged ions
since the radiative decay rates
are large and compete with autoionization rates, i.e. the effect
of dielectronic recombination on electron impact excitation.
We studied in detail the radiation damping effect of dielectronic
recombination, on resonance structures, collision strengths, and
rate coefficients.
We presented the detailed results for O{\sc\,vii} in Delahaye \& Pradhan (2002) 
(figures 3 to 6). We found for this ion that the radiation damping affected
only a small region just below the threshold of convergence. As a consequence
the effective collision strengths $\Upsilon$ are not affected at all. 
However, radiation damping is important for higher-Z elements
since the transition probabilities increase with Z. Pradhan (1983a,b) has
estimated the effect on the z-transition to be 9\% in $\Upsilon$ for Fe{\sc\,xxv}
at the temperature of maximum abundance of helium-like iron, as also confirmed 
by the detailed calculation done by Whiteford et al. 2001.

We calculated the effect of radiation damping and found it to be significant 
to more than 10\% at low temperatures for several transitions between 
levels of \ca19. As the temperature increases the
effect on the rates decreases since the
contribution of the resonances near threshold to the rate becomes less
important.
 
In Figure 3 we present the details of the radiation damping effect on the
collision strengths for the transition  $2 ^3S_1\ -\ 1 ^1S_0$ of \ca19.
We clearly
see that the effect is important in the resonances near threshold and that it
will affect the effective collision strengths at low temperature as we show in
Figure 4.
 
%----------------------------------------------------------------------------- 
\begin{figure} 
%\centering 
\psfig{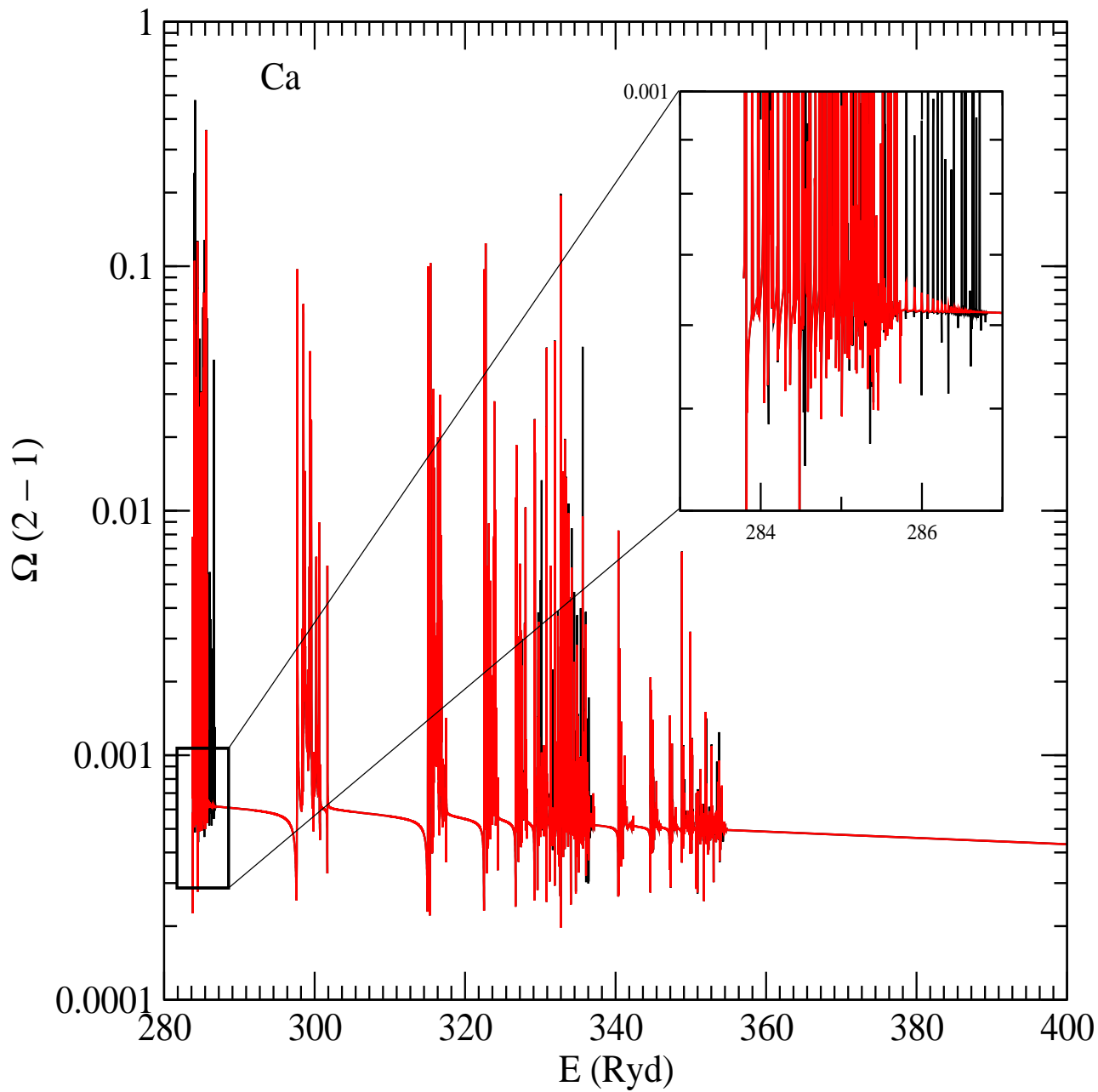} 
\caption{Collision strengths for the z-line (transition $2 ^3S_1\ -\ 1
^1S_0$) in Ca XIX. 
The black line represents the present results when the radiation damping is
neglected and the red one includes the effect. A zoom has been made on the region 
near threshold.} 
\end{figure} 
%---------------------------------------------------------------------------- 

\subsection{Effective Collision Strengths}

 The Maxwellian averaged collision strengths

\begin{equation}
\Upsilon (T) = \int ^{\infty}_{0}
\Omega_{ij}(\epsilon_{j}) e^{-\epsilon_{j}/kT} d(\epsilon_{j}/kT),
\end{equation}
have been computed for all transitions among levels up to the $n=4$.
The collisional rates deduced for levels among $n=4$ might not be as
accurate as for level of lower complexes. Indeed, as has been shown for 
He-like Fe XXV  (Kimura \etal 1999, 2000, and Machado-Pelaez et. al. 2001), 
and for He-like O VII (Delahaye \& Pradhan, 2002)
the resonances arising from the complex $n=N+1$
have a strong effect on transitions to and within the complex $n=N$. The
extension to the $n=5$ complex has been discussed by Whiteford et al. (2001) 
and will not be repeated.

In Figure 4 we present the results of the damping effect on transitions 
($2 ^1P^o_1,~2 ^3S_1,~2 ^3P^o_1,~2^3P^o_2$, corresponding to
the 4 principal lines w, z, x and y, that are of primary
interest in X-ray spectral diagnostics (e.g. Gabriel and Jordan 1969,
Pradhan 1982, Porquet et. al. 2001) . 
We can anticipate that the damping effect will not affect significantly 
the density sensitive ratio R(N$_e$) = z/(x+y) because the 3 transitions 
are affected 
similarly. However the ratio G(T$_e$) = (x+y+z)/w, sensitive to both the
temperature and the ionization balance in the plasma, will be reduced 
due to the unaffected w line compared to the 3 other affected lines. 
Therefore this temperature and ionization diagnostics would be
significantly modified. 

 We also compared the effective collision strengths with previous calculations
from Zhang \& Sampson (1987). The agreement is good with a tendency
towards lower 
values in Zhang \& Sampson and more pronounced for the z line. Indeed, the
difference reaches a maximum of 17\% at low temperatures for the z line, and 
13\% for the w line at high temperature. For the x and y lines the differences
are below 10\% and most of the time within 5\%. 

%----------------------------------------------------------------------------- 
\begin{figure} 
%\centering 
\psfig{figure=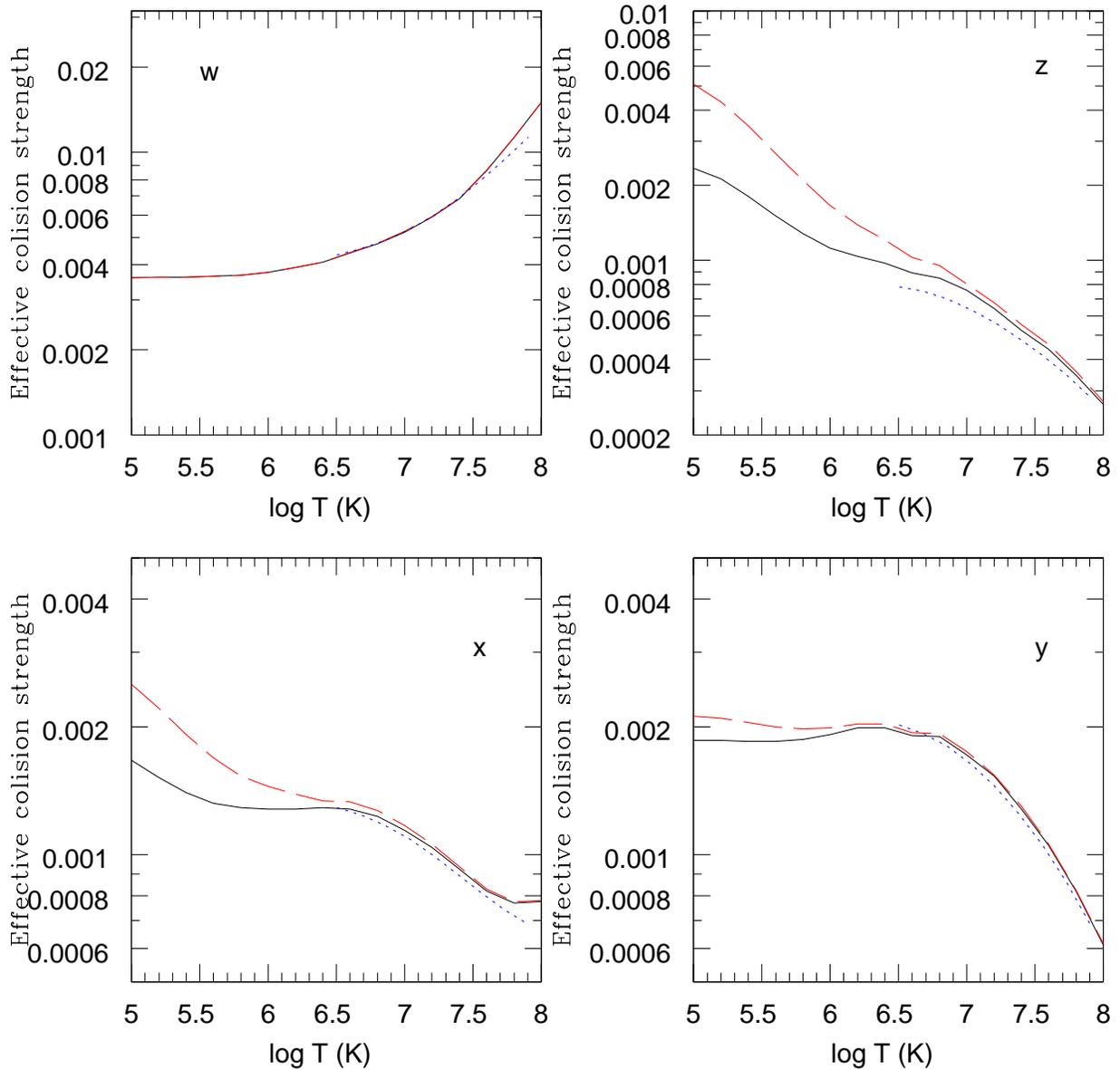,height=17.0cm,width=17.0cm} 
\caption{Effective collision strengths for the principal lines (z, x, y, 
w). 
solid line: Present work with damping,
Dashes: Present work without damping,
dotted line: Zhang \& Sampson 1987.} 
\end{figure} 
%----------------------------------------------------------------------------

From Figure 1 and 2  we can see that the general behavior is as
expected. We can see $\Omega$ decreasing as $Z$ increases. It roughly
follows
\begin{equation}
\Omega \propto (Z+1)^{-2}.
\end{equation}
This comes from the fact that in an isoelectronic sequence the
target is getting more compact as Z increases, the attraction from the highly charged
nucleus being stronger for higher Z.
We also see how the complexes get farther apart as Z increases, decreasing the
interaction between levels of the same symmetry but from different shells.

In order to observe the general behavior along the isoelectronic sequence we
use the reduced effective collision strengths as introduced by
Burgess \& Tully (1992). It allows us to represent on the same plot the effective
collision strengths for each transition for all ions considered in the
sequence in a reduced temperature $T_R$ = T/Z$^3$ domain.

The behavior at high energy normally should follow the
behavior corresponding to the type of transition (Burgess \& Tully 1992).
However, as we increase the charge the spin-orbit coupling is increasing
and as previously shown (Pradhan 1983, also mentioned in Whiteford et al. 2001)
in this case the relativistic effects
are crucial in order to account for this coupling, which of course does not
appear in the LS calculation.
In the present case the coupling between the triplet $P^o$ and the
singlet $P^o$ for example is becoming more and more important as we go to
heavier elements.
In the complex $n=2$, the levels $^1P^o$ and $^3P^o$ are more mixed for heavier
He-like ions.
As a consequence the effective collision strengths for transitions between
the ground state and $2 ^3P^o$ which is an inter-combination transition should
behave at high energy like
\begin{equation}
\Omega \sim const./E^2
\end{equation}
and $\lim_{E\rightarrow \infty} \Omega=0$ but the mixing between the two
levels, singlet and triplet, changes the behavior of this inter-combination
transition by introducing a dipole-allowed component. The
coulomb limit then is not the same any more and does not tend to $0$ when
$E \rightarrow 0$. While Pradhan (1983) showed the trend for He-like Fe, Se
and Mo (Z=26,34 and 42 respectively) the present work demonstrates that the
relativistic effects are important already at lower Z and not only
for species with $Z \ge 20$ (Pradhan 1983). 
In Figure 5 we can see the change
in the behavior of the collision strengths as $E$ increases for this
inter-combination transition between the singlet $S$ and triplet $P^o$.
While He-like N, O and Ne present a discernible trend corresponding to 
inter-combination transitions, it is clearly obvious
in He-like Si, S and Ca.
This is of importance in the
procedure to treat the infinite limit of the collision strengths. Indeed, the
definition of the effective collision strengths (eq. 3.3) supposes an
integration up to infinite energy. But in practice we have to stop at a 
certain point
and make an extrapolation to infinity. As presented
in Burgess and Tully (1992), this extrapolation depends on
the nature of the transition.

%----------------------------------------------------------------------------- 
\begin{figure} 
%\centering 
\psfig{figure=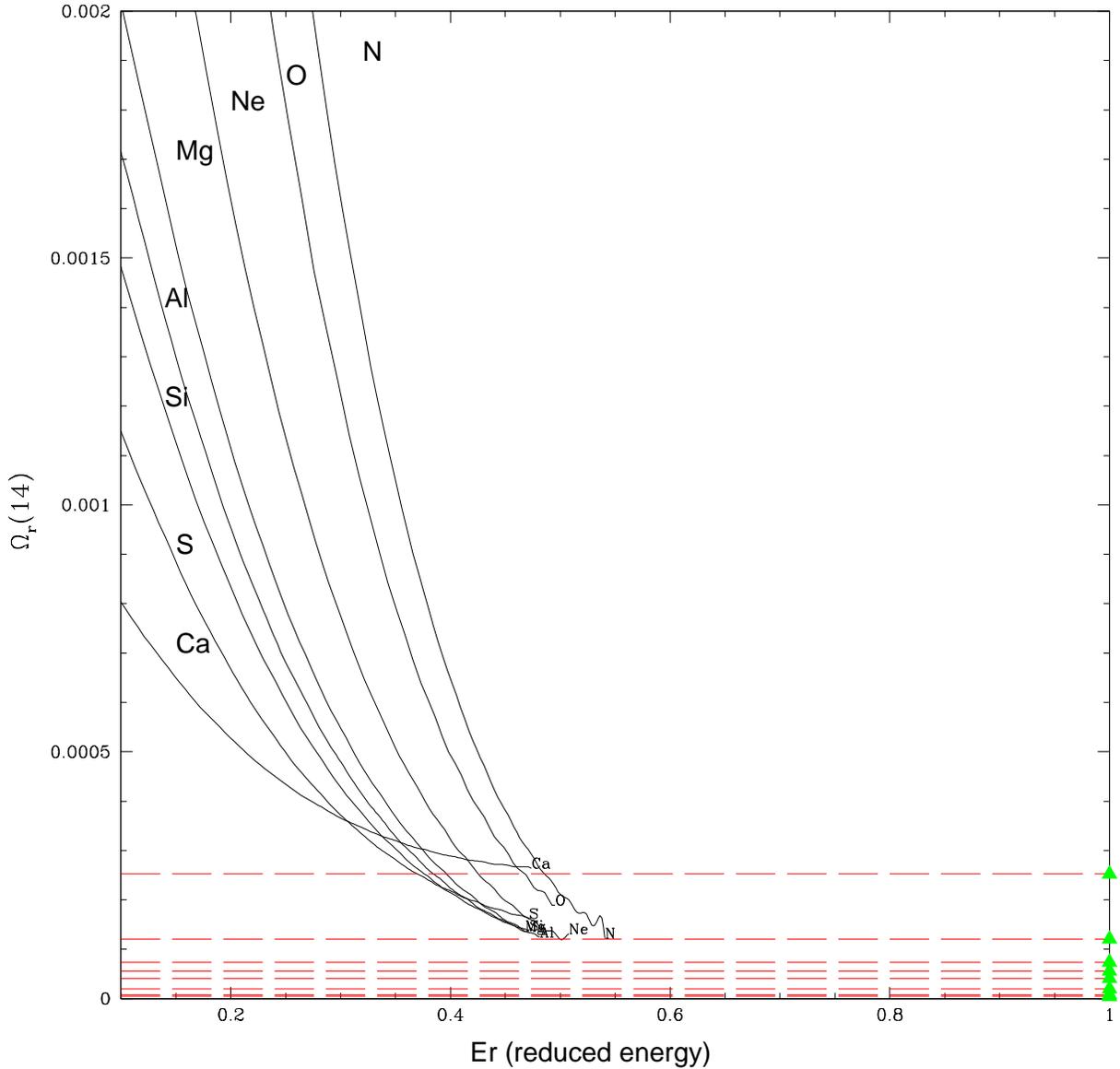,height=17.0cm,width=17.0cm} 
\caption{Detailed of reduced collision strengths for the inter-combination
transition $1^1S_0-2^3P^o_1$. The triangles represent the Coulomb-Born 
limits as $E\rightarrow \infty$.} 
\end{figure} 
%----------------------------------------------------------------------------

In Figure 6 we present a comparison of the effective collision strengths
for the present results with 2 different
treatments for the high energy limit for the transition $1~^1S_0 -
2~^3P^0_1$,
as well as the results from Zhang \& Sampson (1987) (in red).
The solid lines assume a dipole behavior for the inter-comnination transition
while the dotted lines indicate an inter-combination behavior without
the relativistic mixing with the dipole allowed transtion
$1^1S_0-2^1P^o_1$. We can see that the
former treatment is not adapted for the lighter elements and gives an
overestimation of the effective collision strengths at high temperatures but
is necessary for the heavier ones to avoid any underestimation which would
then directly affect the collisional rates.

%----------------------------------------------------------------------------- 
\begin{figure} 
%\centering 
\psfig{figure=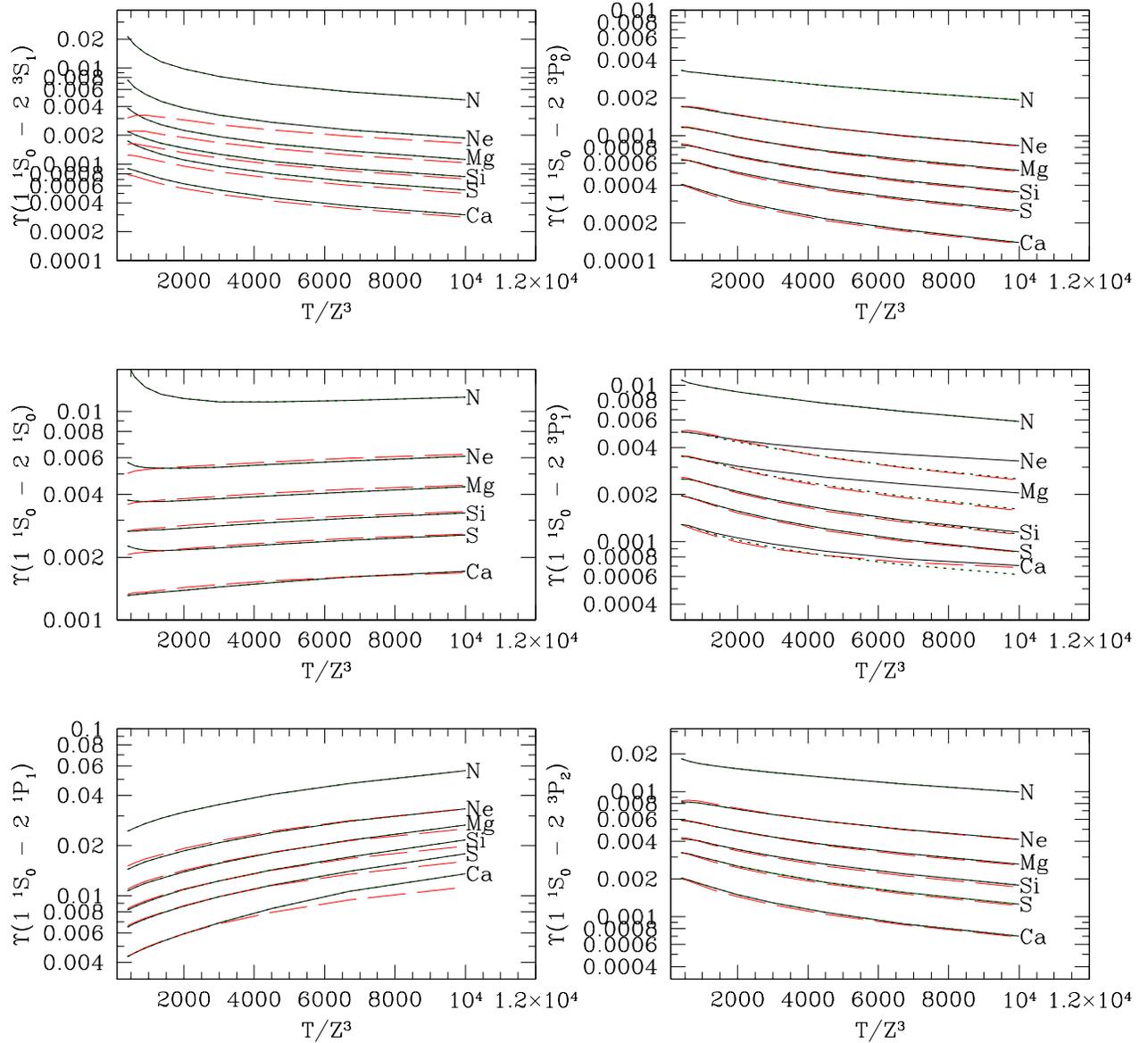,height=17.0cm,width=17.0cm} 
\caption{Effective collision strengths for He-like ions for
transitions among the ground state and $ n=2$ complex.
The solid black line assumes a dipole behavior at high energies for
the inter-combination transition $1~^1S_0~-~2~^3P_1^o$ while the
green dotted line is without relativistic mixing. In red are
the results from Zhang \& Sampson (1987).} 
\end{figure} 
 
%%----------------------------------------------------------------------------- 

\section{Conclusion\protect\\} 

 Some of the general conclusions of the paper are as follows.

 1. The most complete close-coupling calculation using the Breit-Pauli
R-Matrix method has been 
carried out for helium-like ions, including resonances up to $n = 4$
levels.
Detailed studies of radiation damping indicate that it may have a 
significant effect on the detailed collision strengths in a small energy
region below the threshold(s) of convergence, but not on the effective 
collision strengths for the low Z ions of the sequence.
However, radiation damping is important for higher-Z elements 
since the transition probabilities increase with Z and must be taken into
account, especially at low temperature. 

 2. The new results for the important X-Ray line transition may significantly
affect the analysis of X-ray spectra from photoionized sources (e.g. active galactic
nuclei), where ions may be abundant at relatively low temperatures.
In collisional ionized (coronal) sources the new
results may not affect the theoretically computed line intensities
significantly at temperatures close to maximum abundance, but should still
do so at lower temperatures.
It would be preferable to employ the present data in future
collisional-radiative and photoionization models.

 3. As all relevant atomic effects in electron-ion collisions  have been
considered, and resonances have been carefully delineated, we should
expect the present results to be of definitive accuracy. Nonetheless, we
conservatively estimate the precision to be about 10-15\%.

 4. A study of the line ratios R and G in He-like ions is in progress
for diagnostics of astrophysical and laboratory sources.

 5. All data will be electronically available from the first author from
franck.delahaye@obspm.fr.

 The authors would like to thank Dr. Werner Eissner and Dr. Peter Storey for
their help and fuitful discussions. 
This work was supported partially by the NASA Astrophysical Theory Program.
The computational work was carried out at the Ohio Supercomputer Center 
in Columbus, Ohio.

%*** 
%***  E n d   o f   p a g e   1   o f   g a l l e y - m o d e   o u t p u t 
%*** 
 
%------------------------------------------------------------------------ 
 
\section*{References} 
 
\def\amp{{\it Adv. At. Molec. Phys.}\ } 
\def\apj{{\it Astrophys. J.}\ } 
\def\apjs{{\it Astrophys. J. Suppl. Ser.}\ } 
\def\apjl{{\it Astrophys. J. (Letters)}\ } 
\def\aj{{\it Astron. J.}\ } 
\def\aa{{\it Astron. Astrophys.}\ }
\def\aas{{\it Astron. Astrophys. Suppl.}\ } 
\def\aasup{{\it Astron. Astrophys. Suppl.}\ } 
\def\adndt{{\it At. Data Nucl. Data Tables}\ } 
\def\cpc{{\it Comput. Phys. Commun.}\ } 
\def\jqsrt{{\it J. Quant. Spectrosc. Radiat. Transfer}\ } 
\def\jpb{{\it Journal Of Physics B}\ } 
\def\pasp{{\it Pub. Astron. Soc. Pacific}\ } 
\def\mn{{\it Mon. Not. R. astr. Soc.}\ } 
\def\pra{{\it Physical Review A}\ } 
\def\prl{{\it Physical Review Letters}\ } 
\def\zpds{{\it Z. Phys. D Suppl.}\ } 
 
\begin{harvard} 
%\item{}Bell R.H., Seaton M.J., 1985, AdSpR 15, 37 
 
%\item{}Berrington K.A., Burke P.G., Chang J.J., \etal, 1974, Comput. 
%Phys. Commun. 8, 149 
 
%\item{}Berrington K.A., Burke P.G., Le Dourneuf M., \etal, 1978, Comput. 
%Phys. Commun. 14, 367 
 
\item{}Berrington K.A., Eissner, W. and Norrington, P.H. 1995 \cpc 92 290 

\item{}Burgess, A. \& Tully, J.~A. 1992, \aa, 254, 436 
 
%\item{}Burke P.G., Seaton M.J., 1971, Math. Comput. Phys. 10, 1 
 
%\item{}Burke P.G., Hibbert A., Robb W.D., 1971, \jpb 4, 153 
 
\item{}Burke, V. M. and Seaton, M. J. 1986, \jpb 19 L533  

\item{} Delahaye, F., \& Pradhan, A.~K.\ 2002, \jpb, 35, 3377 
 
\item{}Eissner W., Jones M., Nussbaumer H., 1974, Comput. Phys. Commun. 8,270 

\item{} Gabriel, A.~H.~\& Jordan, C.\ 1969, \mn, 145, 241 

\item{} Hummer, D.~G., Berrington, K.~A., Eissner, W., Pradhan, A.~K., Saraph, H.~E., 
 and Tully, J.~A.\ 1993, \aa, 279

\item{}Kaastra J.S., Mewe, R., Liedahl, D.A., Komosa, S., and Brinkman, A.C.
2000 \aa 354 L83

\item{}Kimura,E., Nakazaki, S., Berrington, K.A. and Norrington P.H., 2000, \jpb
17 3449 

\item{}Kimura, E., Nakazaki, S., Eissner, W.~B., and Itikawa, Y., 1999, 
\aasup, 139, 167 

%\item{}Kingston A.E., Tayal S.S., 1983, \jpb 16, 3465 (a)
 
%\item{}Kingston A.E., Tayal S.S., 1983, \jpb 16, L53 (b)

\item{}Machado-Pelaez, M., Mendoza, C. and Eissner, W. (private communication)

\item{} Nussbaumer, H.~\& Storey, P.~J.\ 1978, \aa, 64, 139 

\item{}Porquet D, Mewe R, Dubau J, Raassen A J J, and Kaastra J S
2001 \aa 376 1113
\item{}Porquet D and Dubau J 2000 \aas 143 495

\item{}Pradhan A.K., 1981 \prl 47 79

\item{}Pradhan A.K., 1982 \apj 263 477 

\item{}Pradhan A.K., 1983 \pra 28 2113 (a) 2128 (b)
 
\item{}Pradhan A.K., Seaton M.J., 1985, \jpb 18, 1631 
 
%\item{}Pradhan A.K., Norcross D.W., Hummer D.G., 1981, Ap. J. 246, 1031 
 
%\item{}Pradhan A.K., Norcross D.W., Hummer D.G., 1981, Phys. Rev. A 23,619 
 
\item{}Presnyakov and Urnov A.M., 1979, J. Phys. B 8, 1280 
 
\item{}Sampson D.H., Goett S.J., Clark R.E.H., 1983, Atomic Data Nuclear 
Data Tables 29,467 

%\item{}Steenman-Clark L. And Faucher P., 1984, \jpb  17, 73 

\item{} Whiteford, A.~D., Badnell, N.~R., Ballance, C.~P., O'Mullane,
 M.~G., Summers, H.~P., \& Thomas, A.~L.\ 2001, \jpb, 34, 3179 
 
\item{}Zhang H.L., Sampson D.H., 1987, Ap. J. Supp. Ser. 63, 487 
\end{harvard} 
%------------------------------------------------------------------------ 
 
\enddocument